# Interaction of photons with silver and indium nuclei at energies up to 20 MeV

J. H. Khushvaktov [a,b], M. A. Demichev [a], D. L. Demin [a], S. A. Evseev [a],

M. I. Gostkin [a], V. V. Kobets [a], F. A. Rasulova [a,b,], S. V. Rozov [a], E. T. Ruziev [b],

A. A. Solnyshkin [a], T. N. Tran [a,c], E. A. Yakushev [a], B. S. Yuldashev [a,b]

[a] *Joint Institute for Nuclear Research (JINR), Dubna, Russia*

[b] *Institute of Nuclear Physics of the Academy of Sciences of the Republic of Uzbekistan (INP ASRU), Tashkent, Uzbekistan*

[c] *Institute of Physics of the Vietnam Academy of Science and Technology (IP VAST), Hanoi, Vietnam*

**Abstract**

The yields of photonuclear reactions in the $^{107}$Ag, $^{113}$In, and $^{115}$In nuclei have been measured. Monte Carlo simulations have been performed using the Geant4 code, and the results have been compared with the experimental ones. The isomeric ratios of the yields in the reactions $^{107}$Ag($\gamma$,n)$^{106m,g}$Ag and $^{113}$In($\gamma$,n)$^{112m,g}$In have been determined. The cross sections for the reactions $^{107}$Ag($\gamma$,n)$^{106g}$Ag and $^{107}$Ag($\gamma$,2n)$^{105}$Ag at an energy of 20 MeV have been calculated on the basis of the experimental data.

## 1. Introduction

The study of isomeric states of atomic nuclei provides information on the nature of excited states of atomic nuclei [1-6]. The phenomenon of isomerism of atomic nuclei is associated with a large difference in spins or deformations of the isomeric state relative to the ground state of an atomic nucleus. The probability of isomer population resulting from photonuclear reactions depends on the reaction energy, orbital momentum of emitted particles, momentum and parity of the final state, and probabilities of cascade transitions from higher-lying states [7]. In this work, in addition to the isomeric ratios, the yields and cross sections of photonuclear reactions in silver and indium samples have been investigated.

Natural silver consists of two stable isotopes, with isotope abundance, 51.84 % ($^{107}$Ag) and 48.16 % ($^{109}$Ag). The $^{107}$Ag and $^{109}$Ag nuclei have Z = 47 protons. From the point of view of the shell model, this is a three-proton-hole state with respect to the magic number Z = 50. The number of neutrons in them is N = 60 and 62, respectively. The ground states of $^{107}$Ag ($J^\pi = 1/2^-$) and $^{109}$Ag ($J^\pi = 1/2^-$) are formed as a result of interaction of two fairly complex proton and neutron configurations [8]. Natural indium also consists of two isotopes, a stable one $^{113}$In (4.29 %) and a beta radioactive one $^{115}$In (95.71 %; half-life 4.41·10$^{14}$ years). The number of protons in the odd-even nuclei of $^{113}$In and $^{115}$In is close to the magic number Z = 50, and the number of neutrons is N = 64 and 66, respectively. The spin-parity of the ground state of $^{113}$In and $^{115}$In is determined by the 1g9/2 subshell, which contains nine protons [9].

## 2. Experimental design and data analysis

The experiments have been carried out at the LINAC-200 electron accelerator [10]. A tungsten converter with a size of 4.5 x 4.5 x 0.5 cm was irradiated with 20-MeV electron beams. The samples of silver and indium with respective dimensions of 0.7 x 0.6 x 0.1 cm and 1.0 x 0.7 x 0.1 cm were placed behind the tungsten converter. The $^{nat}$Ag and $^{nat}$In samples were irradiated with a bremsstrahlung flux generated in the tungsten converter during electron beam irradiation with the irradiation time of 40 min, pulse current of 20 mA, pulse frequency of 10 Hz, and pulse duration of 2 μs. Pulse current was measured with an inductive current sensor based on a Rogowski coil. After irradiation, the silver and indium samples were transferred to the test room, and their gamma spectra were measured using an HPGe detector. Eight gamma spectra of each sample were measured with different measurement times. The time from the end of irradiation to the beginning of the measurement of the first spectrum of the $^{nat}$Ag and $^{nat}$In samples was 27 and 36 min, respectively. The gamma spectra obtained were processed using the Deimos32 program [11]. The areas of the identified peaks were determined while the background from the Compton scattering of photons was subtracted. The absolute efficiency of the HPGe detector was measured using standard gamma-ray sources at the same distances from the detector at which the silver and indium samples were examined. The set of gamma-ray sources consisted of $^{54}$Mn, $^{57}$Co, $^{60}$Co, $^{88}$Y, $^{109}$Cd, $^{113}$Sn, $^{133}$Ba, $^{139}$Ce, $^{152}$Eu, $^{228}$Th, and $^{241}$Am. The yields of photonuclear reactions in the samples were determined with the following formula [12]:

$$Y_{exp} = \frac{S_p \cdot C_{abs}}{\varepsilon_p \cdot I_\gamma} \frac{t_{real}}{t_{live}} \frac{1}{N} \frac{1}{N_{e^-}} \frac{e^{\lambda \cdot t_{cool}}}{1 - e^{-\lambda \cdot t_{real}}} \frac{\lambda \cdot t_{irr}}{1 - e^{-\lambda \cdot t_{irr}}}, \qquad (1)$$

where $S_p$ is the full-energy peak area, $\varepsilon_p$ is the full-energy peak detector efficiency, $C_{abs}$ is the self-absorption correction coefficient, $I_\gamma$ is the gamma emission probability, $t_{real}$ and $t_{live}$ are the real time and the live time of the measurement, respectively, $N$ is the number of atoms in a sample, $N_{e^-}$ is the integral number of electrons incident on the tungsten converter, $\lambda$ is the decay constant, $t_{cool}$ is the cooling time, and $t_{irr}$ is the irradiation time.

The values of the parameters of the nuclear reactions studied in this work according to data from [29] and the yields of the measured nuclear reactions are given in the Table 1. $E_{th}$ are reaction thresholds, J, π, and $T_{1/2}$ are the spin, parity, and half-life of the nuclear reaction products, respectively, $E_\gamma$ are the energies of gamma rays emitted by the reaction products, $Y_{exp}$ are the yields of the reactions. The values of the reaction thresholds $E_{th}$ are obtained from the TALYS-1.96 program [17]. For all identified gamma rays of radionuclides that are given in Table 1, the reaction yields were calculated using formula (1). Then, if a radionuclide is identified with more than one gamma line, then the reaction yield values for them are averaged and given in the Table 2.

Table 1. Spectroscopic data [29] on the nuclei products and experimental yields of measured nuclear reactions.

| Nuclear reaction | $E_{th}$ (MeV) | $J^\pi$ of nucleus-product | $T_{1/2}$ | $E_\gamma$ (keV) | $I_\gamma$ (%) | $Y_{exp}$ (atom$^{-1}$ · electron$^{-1}$) |
|---|---|---|---|---|---|---|
| $^{107}$Ag(γ,n)$^{106m}$Ag | 9.55 | 6$^+$ | 8.28 (2) d | 511.842 (28) | 88 (3) | 5.04(31)E-29 |
| | | | | 1045.83 (8) | 29.6 (10) | 4.79(26)E-29 |
| | | | | 717.24 (6) | 28.9 (8) | 5.00(26)E-29 |
| | | | | 450.97 (3) | 28.2 (7) | 4.95(26)E-29 |
| | | | | 616.174 (24) | 21.6 (6) | 4.77(25)E-29 |
| | | | | 748.44 (7) | 20.6 (6) | 4.63(25)E-29 |
| | | | | 1527.65 (19) | 16.3 (13) | 5.15(34)E-29 |
| | | | | 824.79 (15) | 15.3 (4) | 4.89(26)E-29 |
| | | | | 406.17 (3) | 13.4 (4) | 5.15(28)E-29 |
| | | | | 429.64 (5) | 13.2 (4) | 4.93(26)E-29 |
| | | | | 804.34 (13) | 12.4 (5) | 4.82(27)E-29 |
| | | | | 1128.00 (6) | 11.8 (5) | 5.25(30)E-29 |
| | | | | 1199.39 (10) | 11.2 (5) | 5.20(30)E-29 |
| $^{107}$Ag(γ,n)$^{106g}$Ag | 9.55 | 1$^+$ | 23.96 (4) min | 621.94 (3) | 0.316 (8) | 4.99(28)E-27 |
| | | | | 873.48 (4) | 0.199 (5) | 4.76(29)E-27 |
| | | | | 1050.39 (5) | 0.167 (5) | 4.61(32)E-27 |
| | | | | 616.174 (24) | 0.142 (5) | 4.79(32)E-27 |
| | | | | 1128.00 (6) | 0.0721 (25) | 5.18(40)E-27 |
| | | | | 1194.53 (4) | 0.0398 (25) | 4.39(52)E-27 |
| | | | | 1562.24 (5) | 0.0172 (5) | 4.12(70)E-27 |
| $^{107}$Ag(γ,2n)$^{105}$Ag | 17.56 | 1/2$^-$ | 41.29 (7) d | 344.520 (21) | 41 (1) | 2.28(13)E-29 |
| | | | | 280.41 (6) | 30.2 (17) | 2.28(14)E-29 |
| | | | | 63.98 (3) | 10.5 (10) | 2.50(20)E-29 |
| | | | | 443.37 (7) | 10.5 (5) | 1.95(13)E-29 |
| | | | | 319.14 (6) | 4.35 (21) | 2.10(23)E-29 |
| $^{113}$In(γ,γ')$^{113m}$In | 1.45 | 1/2$^-$ | 1.658 (1) h | 391.690 (15) | 64.2 (1) | 2.56(11)E-28 |
| $^{113}$In(γ,n)$^{112m}$In | 9.55 | 4$^+$ | 20.56 (6) min | 156.56 (10) | 13.2 (3) | 1.30(7)E-27 |
| $^{113}$In(γ,n)$^{112g}$In | 9.55 | 1$^+$ | | 617.516 (11) | 4.6 (1) | 4.40(47)E-28 |

| | | | 14.97 (10) min | 606.88 (15) | 1.111 (19) | 4.10(45)E-28 |
|---|---|---|---|---|---|---|
| ¹¹³In(γ,2n)¹¹¹In | 17.23 | 9/2⁺ | 2.8047 (5) d | 245.395 (20) | 94 (1) | 3.19(15)E-29 |
| | | | | 171.28 (3) | 90 (1) | 3.07(15)E-29 |
| ¹¹⁵In(n,γ)¹¹⁶ᵐIn | - | 5⁺ | 54.29 (17) min | 1293.558 (15) | 84.4 (17) | 1.93(8)E-29 |
| | | | | 1097.326 (22) | 56.2 (11) | 2.01(9)E-29 |
| | | | | 416.86 (3) | 27.7 (12) | 2.01(9)E-29 |
| | | | | 2112.312 (22) | 15.5 (4) | 1.97(10)E-29 |
| | | | | 818.718 (21) | 11.5 (4) | 2.04(10)E-29 |
| | | | | 1507.67 (4) | 10.0 (3) | 1.97(10)E-29 |
| ¹¹⁵In(γ,γ')¹¹⁵ᵐIn | 1.27 | 1/2⁻ | 4.486 (4) h | 336.240 (12) | 45.83 (10) | 6.98(25)E-29 |
| ¹¹⁵In(γ,n)¹¹⁴ᵐIn | 9.28 | 5⁺ | 49.51 (1) d | 190.29 (3) | 15.56 (15) | 1.72(6)E-27 |
| | | | | 558.456 (2) | 3.24 (23) | 2.02(10)E-27 |
| | | | | 725.298 (9) | 3.24 (23) | 1.95(10)E-27 |

## 3. Monte Carlo simulation

Currently, Geant4 [13], FLUKA [14], and MCNP6 [15] are the most common Monte Carlo radiation transport codes for photonuclear reaction studies. In this work, we performed simulations with the Geant4 code. To simulate photonuclear interactions in Geant4, the G4PhotoNuclearProcess class [16] is used. Due to the difficulty of using a sufficient number of electrons to determine the number of photonuclear reactions with a small uncertainty, we managed to obtain in Geant4 calculations only fluences of electrons, bremsstrahlung, and neutrons. Further, the yields of the photonuclear reaction and the reaction of neutron capture by the nucleus were determined using formula (2). The reaction cross sections were calculated using the TALYS-1.96 program.

$$Y_{calc} = \int_{E_{thr}}^{E_{max}} f(E)\sigma(E)dE , \qquad (2)$$

where $f(E)$ is the bremsstrahlung fluence (in the case of the neutron capture reaction, the neutron fluence), $\sigma(E)$ is the reaction cross section.

## 4. Results and discussion

Based on the results of processing the measured gamma spectra, we identified photoneutron reactions with the release of up to two neutrons from nuclei, inelastic scattering of photons in nuclei, as well as the reaction of capture of secondary neutrons by the nucleus. Table 2 shows the yields of the above reactions in the ¹⁰⁷Ag,

$^{113}$In, and $^{115}$In nuclei per one electron with an energy of 20 MeV incident on the tungsten converter and their comparison with the simulation results.

Table 2. Experimental yields of measured reactions [atom$^{-1}$ · electron$^{-1}$].

| Reaction | Yield(Exp.) | Calc./Exp. | Reaction | Yield(Exp.) | Calc./Exp. |
|---|---|---|---|---|---|
| $^{107}$Ag(γ,n)$^{106m}$Ag | 4.94(50)E-29 | 1.05(11) | $^{113}$In(γ,n)$^{112g}$In | 4.24(53)E-28 | 1.97(25) |
| $^{107}$Ag(γ,n)$^{106g}$Ag | 4.79(50)E-27 | 0.40(4) | $^{113}$In(γ,2n)$^{111}$In | 3.13(34)E-29 | 1.69(18) |
| $^{107}$Ag(γ,2n)$^{105}$Ag | 2.20(24)E-29 | 1.68(18) | $^{115}$In(n,γ)$^{116m}$In | 1.99(20)E-29 | 0.83(8) |
| $^{113}$In(γ,γ')$^{113m}$In | 2.56(28)E-28 | 0.50(5) | $^{115}$In(γ,γ')$^{115m}$In | 6.98(73)E-29 | 2.51(27) |
| $^{113}$In(γ,n)$^{112m}$In | 1.30(15)E-27 | 1.91(22) | $^{115}$In(γ,n)$^{114m}$In | 1.84(21)E-27 | 1.60(18) |

According to the Nilsson model, the ground state in $^{106}$Ag (1$^+$) can be described as the result of addition of the proton state 1/2$^-$ to the deformed neutron state 1/2$^-$ arising from the splitting of the 1h11/2 shell level. The J$^π$=6$^+$ state can be obtained by adding the proton level 7/2$^+$ to the neutron level 5/2$^+$ formed upon the splitting of the 1g7/2 level in the deformed potential [8]. For the isomeric pair of the odd-odd $^{112}$In nuclei, the spin-parity of the metastable state is J$^π$=4$^+$, and for the ground state, it is J$^π$=1$^+$. One unpaired proton located on the 1g9/2 subshell and a neutron, for example, on the 3s1/2 subshell, apparently form the J$^π$=4$^+$ isomeric state. To form the ground state, the neutron must be on a subshell with a large spin (7/2) in order to form the J$^π$=1$^+$ state together with the proton shell 9/2$^+$ [9].

### 4.1 Isomeric yield ratio

According to the experimental data obtained, we have determined the isomeric ratios of $^{112m}$In/$^{112g}$In and $^{106m}$Ag/$^{106g}$Ag. To determine the isomeric ratios of the yield in the photoneutron reaction $^{113}$In(γ,n)$^{112m,g}$In, the following formula (3) was used:

$$\frac{Y_m}{Y_g} = \left[\frac{\lambda_g(1-e^{-\lambda_m t_{irr}})e^{-\lambda_m t_{cool}}(1-e^{-\lambda_m t_{real}})}{\lambda_m(1-e^{-\lambda_g t_{irr}})e^{-\lambda_g t_{cool}}(1-e^{-\lambda_g t_{real}})}\left(\frac{S_{p,g}}{S_{p,m}}\frac{I_{γ,m}}{I_{γ,g}}\frac{\varepsilon_{p,m}}{\varepsilon_{p,g}}\frac{C_{abs,m}}{C_{abs,g}} - p\frac{\lambda_g}{\lambda_g-\lambda_m}\right) + p\frac{\lambda_m}{\lambda_g-\lambda_m}\right]^{-1}, \quad (3)$$

where $p$ is the branching ratio.

The half-life of the $^{112m}$In (4$^+$) isomeric state is 20.56 min. Emitting a gamma quantum with an energy of 156 keV, it decays to the ground state of $^{112g}$In (1+ ) by internal transition with a branching intensity of 100 %. The gamma quantum yield in the isomeric transition is 13.2 %. The ground state of $^{112g}$In (1$^+$) has a half-life of 14.97 min. It decays via EC to $^{112}$Cd with a probability of 62 % and via β$^-$ to $^{112}$Sn with a probability of 38 %. During the decay of $^{112g}$In (1$^+$), the highest gamma quantum yield is 4.6 %, and the corresponding energy is 617.5 keV. The isomeric ratio Y$_m$/Y$_g$ for the yield of the $^{113}$In(γ,n)$^{112m,g}$In reaction was determined from the area of the peaks of gamma rays with energies of 156.6 and 617.5 keV. The yield ratio $Y_m/Y_g$ is 3.15(36), and it coincides with the literature data [18, 21, 22] as shown in Figure 1a. Using isomeric ratios $Y_m/Y_g$ of $^{112}$In nuclei, the yield of the reaction $^{113}$In(g,n)$^{112g}$In was determined. The yield value of the reaction $^{113}$In(g,n)$^{112m}$In given in Table 1 was determined using formula (1).

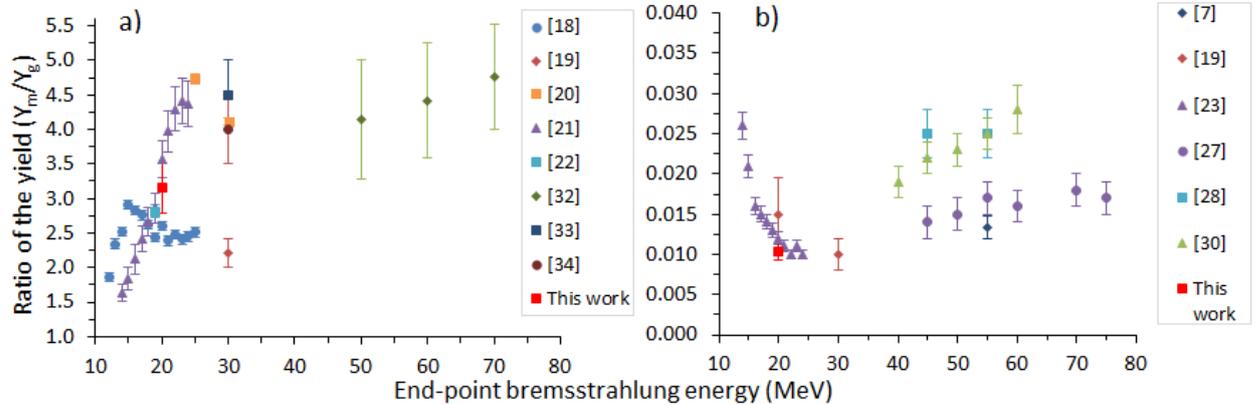

Fig.1. Dependence of the isomeric yield ratio in photoneutron reactions $^{113}$In(γ,n)$^{112m,g}$In (a) and $^{107}$Ag(γ,n)$^{106m,g}$Ag (b) on the end-point bremsstrahlung energy.

In determining the isomeric ratios in the $^{106}$Ag nucleus, the situation was simple. Since the half-lives of the isomeric $J^π=6^+$ ($T_{1/2}$ = 8.28 d) and ground $J^π=1^+$ ($T_{1/2}$ = 23.96 min) states of the $^{106}$Ag nucleus differ, the independent yields $Y_m$ and $Y_g$ were measured. The isomeric ratio for the $^{106}$Ag nucleus is 1.03(11)E-2, which also coincides with the literature data [19, 23] as shown in Figure 1b. It can be seen from the available literature data [7, 18-23, 27-30, 32-34] that an average value of isomeric ratios of $^{112m,g}$In in the reaction $^{113}$In(γ,n) increases with increasing of end-point bremsstrahlung energies (see Fig. 1a). The average value of isomeric ratios of $^{106m,g}$Ag in the reaction $^{107}$Ag(γ,n) in the region of end-point bremsstrahlung energies of 14 to 75 MeV also increases (see Fig. 1b).

### 4.2 Reaction cross section

In this work, although the studies of photonuclear reactions were carried out with bremsstrahlung, it is still possible to determine the cross sections of photonuclear reactions for a specific photon energy using the method of calculating the monochromatic photon spectrum. The reaction cross section can be determined from the reaction yield data by solving formula (2). A method for calculating the quasi-monoenergetic spectrum from the bremsstrahlung spectrum was shown and used in Ref. [24]. This method was also successfully used in Ref. [25]. In this method, in order to determine reaction cross sections, it is required to calculate the effective photon spectrum, the shape of which is close to the spectrum of monoenergetic photons. To calculate the monoenergetic spectrum of photons, we used the calculated bremsstrahlung spectra for electrons with energies of 19, 20, and

21 MeV. When calculating the bremsstrahlung spectra with the Geant4 code, we used statistics of two billion electrons. The bremsstrahlung spectra of the tungsten converter for electrons with energies of 19, 20 and 21 MeV and the calculated quasi-monoenergetic spectrum of photons with an energy of 20 MeV are shown in Figures 2a and 2b, respectively.

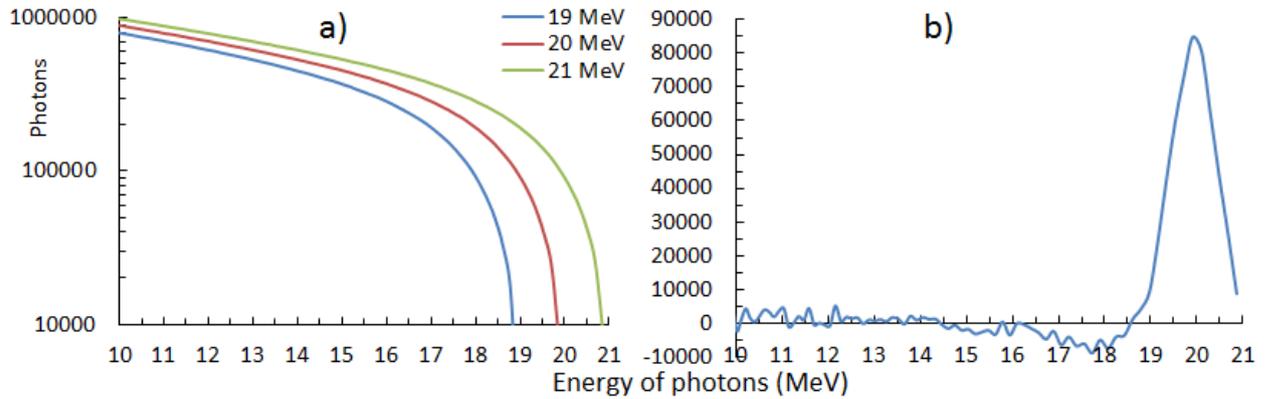

Fig.2. Bremsstrahlung spectra of tungsten converter for electrons with energies of 19, 20 and 21 MeV (a). Quasi-monoenergetic spectrum of 20 MeV photons (b).

To determine the cross section of the reaction, it is also required to determine the value of the yield of the reaction caused by photons with energies of 19 to 21 MeV using the experimental data. To do this, the experimental and calculated values of the yield of the reaction with bremsstrahlung produced by electrons with energies of 19, 20, and 21 MeV are needed. We have the experimental value of the reaction yield only for electrons with an energy of 20 MeV. For 19- and 21-MeV electrons, they were corrected from the calculated data, by considering the ratios as $\frac{Y_{calc}(19\ MeV)}{Y_{calc}(20\ MeV)} = \frac{Y_{exp}(19\ MeV)}{Y_{exp}(20\ MeV)}$ and $\frac{Y_{calc}(21\ MeV)}{Y_{calc}(20\ MeV)} = \frac{Y_{exp}(21\ MeV)}{Y_{exp}(20\ MeV)}$ to be correct. The cross sections for photoneutron reactions $^{107}$Ag($\gamma$,n)$^{106g}$Ag and $^{107}$Ag($\gamma$,2n)$^{105}$Ag at an energy of 20 MeV determined by the above method are 19.0 ± 5.7 and 29.4 ± 10.0 mb, respectively, and coincide with the literature data [26, 35, 36] measured with mono-energetic photons, as shown in Figures 3a and 3b. The values of the flux-weighted average cross-sections at bremsstrahlung end-point energies from the literature data [31] are also shown in Figures 3a and 3b. The region of giant dipole resonance (9-21 MeV), in which only one nucleon is excited when a $\gamma$-quantum is absorbed by the nucleus, can be seen in Figure 3a. Above this region, the quasideuteron mechanism of photoabsorption begins to dominate, in which photons begin to be absorbed predominantly by quasideuterons.

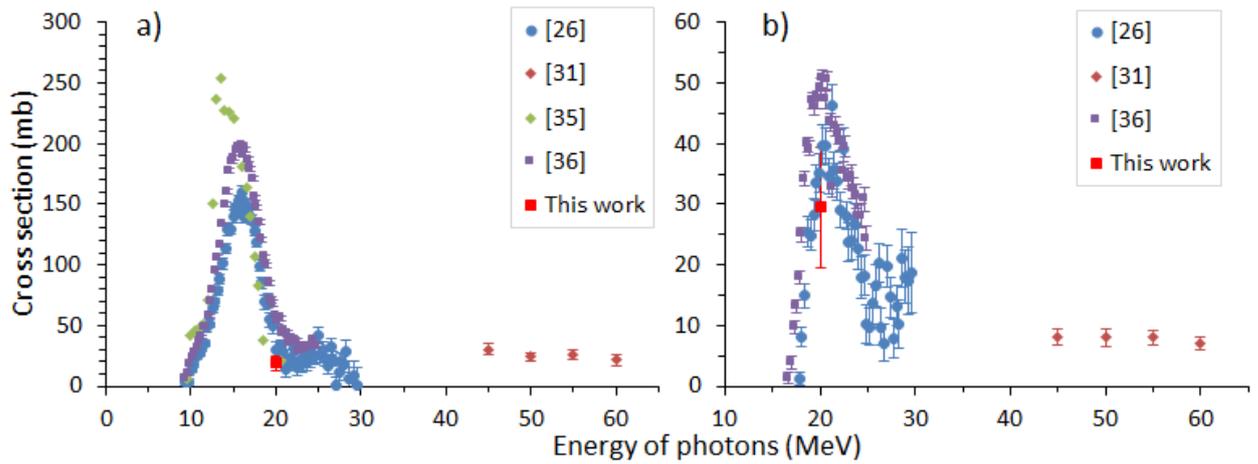

Fig.3. Cross sections for photoneutron reactions $^{107}$Ag($\gamma$,n)$^{106g}$Ag (a) and $^{107}$Ag($\gamma$,2n)$^{105}$Ag (b).

## 5. Conclusion

The interaction of bremsstrahlung with silver and indium nuclei at an accelerated electron energy of 20 MeV has been studied experimentally and theoretically. The yields of photonuclear reactions in $^{107}$Ag, $^{113}$In, and $^{115}$In nuclei were measured. Also, the yield of the capture reaction of secondary neutrons with the $^{115}$In (9/2$^+$) nucleus with the formation of the $^{116m}$In (5$^+$) isomeric state was measured. Monte Carlo simulations were performed using the Geant4 code, and the results were compared with the experimental ones. According to the comparison results, the ratio Calc./Exp. is in the range of 0.40-2.51. The isomeric ratios of the yields in the reactions $^{107}$Ag($\gamma$,n)$^{106m,g}$Ag and $^{113}$In($\gamma$,n)$^{112m,g}$In were determined, and they coincide with the literature data. On the basis of the experimental data, the cross sections for the reactions $^{107}$Ag($\gamma$,n)$^{106g}$Ag and $^{107}$Ag($\gamma$,2n)$^{105}$Ag at an energy of 20 MeV were determined, and these values also coincide with the literature data.

### Acknowledgments

The authors express their gratitude to the team of the LINAC-200 electron accelerator and the management of the Joint Institute for Nuclear Research for their support in conducting the experiments.